\documentclass[prd,showpacs,amsmath,amssymb,superscriptaddress,preprintnumbers,nofootinbib,showkeys]{revtex4}
\usepackage{setspace}
\begin{document}

\newcommand{\nablab}{{\mathop {\rule{0pt}{0pt}{\nabla}}\limits^{\bot}}\rule{0pt}{0pt}}

\title{Kinetics of relativistic axionically active plasma in the field of dynamic aether. \\ Part I: General formalism and
new concept of equilibrium states}

\author{Alexander B. Balakin and Kamil R. Valiullin}
\email{Alexander.Balakin@kpfu.ru}
\email{kr_valiullin@mail.ru} 
\affiliation{Department of
General Relativity and Gravitation, Institute of Physics, Kazan
Federal University, Kremlevskaya str. 18, Kazan 420008,
Russia}

\date{\today}

\begin{abstract}
We establish an extended version of the kinetic theory of the relativistic axionically active multi-component plasma, which is based on the inclusion of a unit time-like vector field, associated with the velocity of dynamic aether, into the scheme of interactions.
The proposed extension of the plasma theory can be indicated as semi-phenomenological. This term means that master equations for the gravitational, electromagnetic, pseudoscalar (axionic) and vector (aetheric) fields  are derived using the Lagrange formalism, the kinetic equations for the distribution functions are obtained by the methods of the covariant statistics, however, the relativistic forces acting from the listed fields on the plasma particles are reconstructed phenomenologically based on classical analogies. We consider this work as a first part of a trilogy; here we presented, first, the complete formalism of this extended theory, second, the classification of forces, third, the new concept of equilibrium states in the relativistic axionically - aetherically active plasma.
\end{abstract}
\maketitle

\section{Introduction}

The covariant relativistic version of the kinetic theory was created at the turn of the 50-60s of the last century.
The history of this event is described in detail in the book \cite{RKT}. The theory of relativistic plasma built on the base of this kinetic theory had enormous success in many fields of science, especially in astrophysics and cosmology (see, e.g., \cite{P1,P2,P3}). Two major tasks can be identified in this context. The first one is the investigations of equilibrium plasma configurations in the vicinity of static astrophysical objects; plasma was considered in the regime of frequent collisions. The second problem was associated with the electromagnetic waves propagation in the relativistic plasma; the Vlasov theory of cooperative electromagnetic field in the collisionless plasma was the appropriate model for such studies.
The canonic plasma theory is based on the Faraday - Maxwell version of classical electrodynamics \cite{Jackson,Hehl}. New trend in the plasma theory appears, when the axion electrodynamics was established. The term axion electrodynamics was first mentioned in 1983 \cite{Sikivie} and has been widely used since then (see, e.g., \cite{Wilczek1987}). This was preceded by the formulation of a recipe for CP invariance conservation \cite{PQ}, and the prediction of the existence of a new massive pseudo-Goldstone boson \cite{W1,W2}, subsequently called an axion. We proposed to use the term axionically active plasma in \cite{B0} by analogy with the term magneto-active plasma in the canonic theory; some problems of such a theory are formulated and solved in the works \cite{B1,B2,B3}.

A new page in the theory of relativistic plasma was opened in connection with the emergence of the theory of dynamic aether \cite{J1,J2,J3}. We plan to study the following scheme of interaction of the axionically active plasma with the vector field, associated with the velocity of aether. First, we assume that there exists a direct channel of influence of the aether on the plasma particles. The basic element in this context is a set of forces, which contain the aether velocity itself and its covariant derivatives; these forces are incorporated into the kinetic equations. Also, there are two indirect channels of coupling via the aetherically induced changes in the structure of the electromagnetic and axion fields, respectively. Mathematical description of the formalism of coupling between photons and aether is prepared in \cite{BaLe}. In the work \cite{AA} the theory of interaction between aether and axion field is elaborated.

What do we mean speaking about a new concept of equilibrium states in the plasma interacting with pseudoscalar (axion) and unit vector (aether) fields? In the canonic plasma theory the macroscopic velocity of plasma motion appears as the result of normalization of the time-like Killing vector, if the space-time admits such Killing vector. The norm of the corresponding Killing vector describes the distribution of the plasma temperature. In other words, the basic characteristics of the  plasma in the canonic equilibrium state are predetermined by the space-time symmetry, or more precisely, by the properties of the gravitational field. When we deal with the dynamic aether, the time-like unit vector already exists, and we assume that namely the aether predetermines the properties of the equilibrium characteristics of the plasma. Clearly, in order to provide such new equilibrium state the aether has to act on the plasma particles by some specific forces. One of such forces is shown to be an analog of the classical Stokes friction force, which acts on the macroscopic particle in the flow of viscous fluid. Another one is the analog of the tidal force. Modeling of cosmic (anti)friction forces without axion and aether fields has been fulfilled in the works \cite{anti1,anti2,anti3,anti4}. Now we extended the list of the corresponding forces by including the aether velocity vector into the list of players.

The development of the theory of the aetherically-axionically active plasma is the aim of our investigations. The presented work is the first part of a trilogy: here we elaborated the formalism of this theory, we obtained the total set of master equations and formulated the concept of equilibrium in the aetherically-axionically active plasma. In the second part of work we plan to solve equations of the plasma particles dynamics under the influence of various forces for the cosmological space-time platform. In the third part of the work we plan to focus on the problem of dispersion relations for electromagnetic waves propagating in the aetherically-axionically active plasma.

\section{The formalism}

\subsection{Semi-phenomenological approach to the description of the system evolution}

We use the semi-phenomenological approach, which is based on the synthesis of the relativistic statistics for description of the plasma kinetics, and of the Lagrange approach for the description of fields, which act on the plasma particles directly or in an indirect way. We reconstruct the forces, which act on the plasma particles, using three fields: first, the unit vector field $U^j$, associated with the aether velocity four-vector, second, the pseudoscalar field $\phi$ describing the axionic dark matter, third, the electromagnetic field with the potential $A_j$. Gravitational field is considered to be responsible for the space-time platform, on which the plasma processes take place. The total action functional is of the form
\begin{equation}
	{-}S_{(\rm total)} {=} \int d^4x \sqrt{{-}g} \left\{ L_{(\rm EA)} {+} L_{(\rm A)} {+} L_{(\rm EM)} {+} L_{(\rm P)} \right\} \,.
	\label{0}
	\end{equation}
The first element in this integral relates to the Einstein-aether Lagrangian
\begin{equation}
	L_{(\rm EA)} = \frac{1}{2\kappa}\left[R{+}2\Lambda {+} \lambda\left(g_{mn}U^mU^n {-} 1 \right) {+}
	K^{ab}_{\ \ mn} \nabla_a U^m  \nabla_b U^n \right] \,,
	\label{01}
	\end{equation}
in which $\kappa$ is the Einstein constant, $\Lambda$ is the cosmological constant, $R$ is the Ricci scalar, $\lambda$ is the Lagrange multiplier. The Jacobson constitutive tensor
\begin{equation}
	K^{ab}_{\ \ mn} = C_1 g^{ab}g_{mn} {+} C_2 \delta^{a}_{m} \delta^{b}_{n} {+}C_3 \delta^{a}_{n} \delta^{b}_{m} {+} C_4 U^a U^b g_{mn}
	\label{11}
	\end{equation}
contains four dimensionless coupling constants $C_1$, $C_2$, $C_3$, $C_4$.
Variation of the total action functional with respect to $\lambda$ gives the normalization condition $U_j U^j =  1$. Also, we introduce the aetheric effective metric
\begin{equation}
G^{mn} = g^{mn} + {\cal A} U^m U^n \,,
\label{G}
	\end{equation}
and consider ${\cal A}$ as an auxiliary guiding parameter. We have to mention that in this work ${\cal A}$ is considered to be constant, but in principle, it can be a guiding function, depending on four differential scalars based on irreducible representation of the covariant derivative $\nabla_m U_n$ (see, e.g., \cite{SH1,SH2,SH3} for details).

The Lagrangian of the dimensionless pseudoscalar (axion) field is presented as follows:
\begin{equation}
	L_{(\rm A)} = \frac12 \Psi^2_0   \left[V(\phi) {-} \left(g^{mn} + {\cal A} U^m U^n \right)\nabla_m \phi \nabla_n \phi \right] \,,
	\label{132}
	\end{equation}
where the potential of the axion field is chosen in the periodic form
	\begin{equation}
	V(\phi) = 2 m^2_A \left[1- \cos{\phi} \right] \,.
	\label{13}
	\end{equation}
Here $m_A$ is the axion mass, and the constant $\Psi_0$ is reciprocal to the constant of the axion - photon coupling.

The Lagrangian of the electromagnetic field has the standard form for linear electrodynamics
\begin{equation}
	{\cal L}_{(\rm EM)} = \frac14 C^{mnpq}F_{mn} F_{pq} \,,
	\label{73}
	\end{equation}
where $F_{mn}$ is the Maxwell tensor, $F_{mn} = \nabla_m A_n {-} \nabla_n A_m$. The Tamm constitutive tensor  	
	\begin{equation}
	C^{mnpq} = \frac12 \left[\left(g^{mp} g^{nq}{-}g^{mq}g^{np} \right) {+}  \sin{\phi} \epsilon^{mnpq} {+}
	 {\cal A}\left(g^{mp}U^n U^q {-} g^{mq}U^n U^p {+} g^{nq}U^m U^p {-} g^{np}U^m U^q  \right) \right]
	\label{74}
	\end{equation}
contains two important details. First, the term with the Levi-Civita (pseudo)tensor $\epsilon^{mnpq}$ describes the axion - photon coupling; the product $\sin{\phi} \epsilon^{mnpq}$ forms the true tensor; for small $\phi$ the corresponding term in the Lagrangian takes the classical form $\frac14 \phi F^{*mn} F_{mn}$ \cite{WTN} with the dual Maxwell tensor $F^{*mn} \equiv \frac12 \epsilon^{mnpq} F_{pq}$.
Second, the guiding parameter ${\cal A}$ in the aetherically extended electrodynamics plays the same role as the square of the effective refraction index $n_{eff}$ in the standard Faraday-Maxwell theory; ${\cal A} = n^2_{eff}-1$. In this sense the Tamm constitutive tensor (\ref{74}) describes a spatially isotropic homogeneous transparent medium, which moves with the velocity four-vector $U^j$. This analogy emphasizes that the dynamic aether interacting with the electromagnetic field behaves as an effective medium with the magnetic impermeability $\frac{1}{\mu}=1$ and dielectric permittivity $\varepsilon = 1+ {\cal A}$.

The Lagrangian of the plasma $L_{(\rm P)}$ is not presented in explicit form, and we use an alternative way for description of this subsystem. In fact, we postulate the structure of relativistic kinetic equations and use the macroscopic moments of the distribution functions in order to link the results of the Lagrange approach with the results of a statistical approach.

\subsection{Field equations}

\subsubsection{Master equations of electrodynamics}
In order to explain the idea of semi-phenomenological approach, which  we use, let us start with derivation of the electrodynamic equations.
Variation of the action functional with respect to the potential of the electromagnetic field yields 	
\begin{equation}
	\nabla_n \left( C^{jnpq} F_{pq} \right) = - \frac{\delta}{\delta A_j} L_{(\rm P)} \,.
\label{T01}
	\end{equation}
From the physical point of view the variational derivative in the right-hand side of (\ref{T01}) is proportional to the total electric current generated by plasma particles. This means that we can make the following replacement:
\begin{equation}
- \frac{\delta}{\delta A_j} L_{(\rm P)} \ \Rightarrow \ - 4\pi \sum_{(a)} e_{(a)} N^j_{(a)} \,,
\label{T021}
	\end{equation}
where $N^j_{(a)}$ is the time-like four-vector describing the flow of particles of the sort $(a)$ possessing the electric charge $e_{(a)}$. The corresponding scalar of the particles number density is defined as ${\cal N}_{(a)} \equiv \sqrt{g_{jk} N^j_{(a)}N^k_{(a)}}$. Below we will link this four-vector with the first moment of the distribution function.
Also, we have to add the set of equations $\nabla_k F^{*ik} =0 $,
which converts into identity, when we replace $F_{mn}$ with $\nabla_m A_n {-}\nabla_n A_m$. Thus, the set of electrodynamic equations reads
\begin{equation}
	\nabla_n \left[F^{jn} + \sin{\phi} F^{*jn} + {\cal A}U_q \left(F^{jq}U^n - F^{nq}U^j \right)  \right] = - 4\pi \sum_{(a)} e_{(a)} N^j_{(a)} \,, \quad \nabla_k F^{*ik} =0  \,.
	\label{T1}
	\end{equation}
	
\subsubsection{Extended Jacobson's equations}

Variation of the total action functional with respect to the vector field $U^j$ gives the balance equation
\begin{equation}
	\nabla_a {\cal J}^{a}_{\ j}  = \lambda  U_j  {+}  C_4 U^s \nabla_s U_m \nabla_j U^m  {+}
 \kappa {\cal A} F_{mj} F^{mq} U_q    - \kappa {\cal A}\Psi^2_0  U^s (\nabla_s \phi) (\nabla_j \phi) + \kappa \frac{\delta L_{(\rm P)}}{\delta U^j} \,.
	\label{AEq}
	\end{equation}
The new tensor ${\cal J}^{a}_{ \ j}$, which is defined using the constitutive law 		
\begin{equation}
		{\cal J}^{a}_{ \ j} = K^{ab}_{\ \ jn} \nabla_b U^n \,,
	\label{16}
	\end{equation}	
 can be indicated as Jacobson's tensor. Convolution of (\ref{AEq}) with the aether velocity four-vector $U^j$ gives the
 Lagrange multiplier
\begin{equation}
	\lambda =  U^j \nabla_a {\cal J}^{a}_{\ j}  {-} C_4 DU_m DU^m  {+}{\cal A}\left[ \kappa \Psi^2_0  (D \phi)^2 {-} \kappa  F_{mj}U^j F^{mq} U_q  \right] - \kappa U^j \frac{\delta L_{(\rm P)}}{\delta U^j} \,,
	\label{150}
	\end{equation}
where we introduced the operator of convective derivative $D \equiv U^s \nabla_s$.	The four-vector $\frac{\delta L_{(\rm P)}}{\delta U^j}$ has to be modeled phenomenologically using the macroscopic moments of the distribution functions.

\subsubsection{Master equations of the axion field}

Variation of the action functional with respect to the pseudoscalar field gives the master equation for the axion field
\begin{equation}
	\nabla_m \left[\left(g^{mn} {+} {\cal A} U^m U^n \right)\nabla_n \phi \right] {+} m^2_A  \sin{\phi}  = {-} \frac{1}{4\Psi_0^2} \cos{\phi} F^*_{mn}F^{mn} {-} \frac{1}{\Psi_0^2} \frac{\delta L_{(\rm P)}}{\delta \phi} \,.
	\label{24}
	\end{equation}
Again, the pseudoscalar $\frac{\delta L_{(\rm P)}}{\delta \phi}$ has to be modeled phenomenologically using the macroscopic moments of the distribution functions.	

\subsubsection{Master equations for the gravitational field}

Variation of the action functional with respect to metric gives the equation of the gravity field
\begin{equation}
	R_{ik} - \frac12 R g_{ik} - \Lambda g_{ik} = T^{(\rm U)}_{ik} + \kappa T^{(\rm A)}_{ik} + \kappa T^{(\rm EM)}_{ik} + \kappa T^{(\rm P)}_{ik}\,.
	\label{25}
	\end{equation}
The stress-energy tensors of the unit vector field, of the axion field, of the electromagnetic field are known to have the following forms, respectively: 	
	\begin{equation}
	T^{(\rm U)}_{ik} =
	\frac12 g_{ik} \ K^{abmn} \nabla_a U_m \nabla_b U_n {+}
	\label{326}
	\end{equation}
	$$
	{+}\nabla^m \left[U_{(i}{\cal J}_{k)m} {-}
	{\cal J}_{m(i}U_{k)} {-}
	{\cal J}_{(ik)} U_m\right]{+} U_iU_k U_j \nabla_a {\cal J}^{aj} {+}
	$$
	$$
	{+}C_1\left[(\nabla_mU_i)(\nabla^m U_k) {-}
	(\nabla_i U_m )(\nabla_k U^m) \right] {+} C_4 \left( D U_i D U_k {-} U_iU_k DU_m DU^m \right)\,,
	$$
	
	\begin{equation}
	T^{(\rm A)}_{ik}= \Psi^2_0 \left[\nabla_i \phi \nabla_k \phi  {+} {\cal A} (D \phi)^2 \left(U_iU_k {-}\frac12 g_{ik}\right) +
	\frac12 g_{ik}\left(V {-} \nabla_s \phi \nabla^s \phi  \right) \right]\,,
	\label{T76}
	\end{equation}
	
	\begin{equation}
	T^{(\rm EM)}_{ik}=  \left( \frac14 g_{ik} F_{mn} F^{mn} - F_{in} F_{k}^{\ n} \right) +
	{\cal A}\left[\left(\frac12 g_{ik}- U_i U_k \right)E_m E^m - E_i E_k  \right]\,.
	\label{T91}
	\end{equation}
For simplicity, we introduced here the electric four-vectors $E^j = F^{jk}U_k$.	
As for the stress-energy tensor associated with plasma particles, we follow the standard rule that it is proportional to the second order macroscopic moment of the distribution functions.	

\subsection{Covariant relativistic kinetic equations and macroscopic moments}

The formalism of kinetic theory is based on a set of semi-phenomenologically derived relativistic kinetic equations
\begin{equation}
\frac{p^j}{m_{(a)} c} \left(\frac{\partial }{\partial x^j} - \Gamma^l_{jk} p^k \frac{\partial }{\partial p^l}\right)f_{(a)} + \frac{\partial}{\partial p^i} \left[{\cal F}_{(a)}^i f_{(a)} \right] = \sum_{(b)} {\cal J}_{(a)(b)} \,.
\label{KE1}
\end{equation}
The set of distribution functions $f_{(a)}(x^j, p^k)$ is marked by the index of particle sort  $(a)$; they are functions of coordinates $x^j$ and of the particle momenta $p^k$, which have the status of random variables. The terms ${\cal F}_{(a)}^i$ relate to the force four-vectors, which act on the particles of the sort $(a)$.  The object ${\cal J}_{(a)(b)}$ describes the collision integral of the particles, which are associated with the sorts $(a)$ and $(b)$.
Integration in the Phase Space with the measure $dP$ gives the set of macroscopic moments
\begin{equation}
{\cal M}_{(a)}^{j_1...j_s} \equiv \int dP f_{(a)} p^{j_1} \cdot \cdot \cdot p^{j_s} \,, \quad dP = d^4p \sqrt{-g} \ \delta\left[p^lp_l- m^2_{(a)}c^2\right] \ \eta(p^l U_l) \,.
\label{KE2}
\end{equation}
Delta function in this measure guarantees that the momentum four-vector of the particle with the mass $m_{(a)}$ is normalized. The Heaviside function $\eta$ rejects the contributions with negative energy $p^l U_l<0$.
Two moments enter into the classical formalism of the kinetic theory: the four-vector of particle number of the chosen sort  $N^j_{(a)}$, and the corresponding part of the stress-energy tensor $T^{ls}_{(a)}$
\begin{equation}
N^j_{(a)} = \frac{1}{m_{(a)} c} \int dP \ p^j f_{(a)} \,, \quad T^{ls}_{(a)} = \frac{1}{m_{(a)}} \int dP \ p^l p^s f_{(a)} \,.
\label{KE3}
\end{equation}
The choice of coefficients in front of integrals relates to the assumptions that $N^j_{(a)}$ has the dimensionality of particle number density, and  $T^{ls}_{(a)}$ has the dimensionality of energy density.

We follow the classical idea that for the elastic collisions, for which $\sum_{(b)} \int dP {\cal J}_{(a)(b)} =0$,  the first moment satisfies the balance equation
\begin{equation}
\nabla_j N^j_{(a)} = - \int dP \frac{\partial}{\partial p^j} \left[{\cal F}_{(a)}^j f_{(a)} \right] =  0 \,.
\label{KE4}
\end{equation}
As usual, $\nabla_j$ is the covariant derivative.
In other words, we assume that the particle number conservation law takes place for arbitrary set of forces ${\cal F}_{(a)}^i$.
Similarly, due to the relation $\sum_{(a)(b)} \int dP p^j {\cal J}_{(a)(b)} =0$, we obtain the known balance equations, which include the total stress-energy tensor
\begin{equation}
\nabla_j \sum_{(a)}T^{lj}_{(a)} = \sum_{(a)} c \int dP f_{(a)} {\cal F}_{(a)}^l \,.
\label{KE5}
\end{equation}

\subsection{Phenomenological modeling of the sources appeared in the field equations}

As it was shown above, in the electrodynamic equations (\ref{T01}) the new source term appears, which has the form
$\frac{\delta}{\delta A_j} L_{(\rm P)}$. Keeping in mind that it should be expressed via the electric current, one can state that
\begin{equation}
- \frac{\delta}{\delta A_j} L_{(\rm P)} = - 4\pi \sum_{(a)} e_{(a)} N^j_{(a)} = - 4\pi \sum_{(a)}  \frac{e_{(a)}}{m_{(a)} c} \int dP \ p^j f_{(a)}  \,.
\label{T041}
	\end{equation}
In other words, this source term can be identified using the first integrals of the distribution functions.
Similarly, the vectorial term  $\kappa \frac{\delta}{\delta U^j} L_{(\rm P)}$ appears in the modified Jacobson's equation (\ref{AEq}).
Our ansatz is that it can be expressed as a linear combination of the first and second moments of the distribution functions as follows:
\begin{equation}
\kappa \frac{\delta}{\delta U^j} L_{(\rm P)} = \sum_{(a)} \left[\omega_{(a)} N_{(a) j} + \Omega_{(a)} T_{(a) jk} U^k \right]  \,.
\label{T031}
	\end{equation}
Two sets of parameters $\omega_{(a)}$ and $\Omega_{(a)}$ are the subject of phenomenological modeling.
The pseudoscalar term ${-} \frac{1}{\Psi_0^2} \frac{\delta L_{(\rm P)}}{\delta \phi}$
appeared in (\ref{24}), can be also expressed as a linear combination of the first and second macroscopic moments
\begin{equation}
{-} \frac{1}{\Psi_0^2} \frac{\delta L_{(\rm P)}}{\delta \phi} = \sin{\phi} \sum_{(a)} \left[\alpha_{(a)} N_{(a) j} U^j + \beta_{(a)} T_{(a) jk} g^{jk} + \gamma_{(a)} T_{(a) jk} U^j U^k \right]  \,.
\label{Tr031}
	\end{equation}
In this decomposition three sets of phenomenological parameters appear.
Finally, the term $T^{(\rm P)}_{ik}$ appeared in (\ref{25}) is the sum of the stress-energy tensors
\begin{equation}
T^{(\rm P)}_{ik} = \sum_{(a)} T_{(a) ik} \,.
\label{T051}
	\end{equation}

\subsection{Classification of forces}

Characteristic equations associated with the kinetic equation (\ref{KE1}) can be standardly written in the form
\begin{equation}
\frac{{\cal D} p^j}{ds} = {\cal F}^j_{(a)} \,, \quad \frac{{\cal D} p^j}{ds} \equiv \frac{d p^j}{ds} + \Gamma^j_{lk} p^l \frac{p^k}{m_{(a)}c} \,.
\label{F0}
\end{equation}
We follow the general idea that the rest mass of particles $m_{(a)}$ conserves, so that
\begin{equation}
p_j \frac{{\cal D} p^j}{ds} = \frac12 \frac{{\cal D} (p_jp^j)}{ds} = \frac12 \frac{d [m^2_{(a)}c^2] }{ds} =0 \,.
\label{Fr0}
\end{equation}
This means that the force is orthogonal to the particle momentum four-vector
\begin{equation}
p_j {\cal F}^j_{(a)} =0 \,.
\label{F1}
\end{equation}
There are two specific constructions providing the requirement (\ref{F1}). The first one (basic) contains the projector with respect to the particle momentum, and the force has the following general form with arbitrary vectorial objects ${\cal W}^k_{(a)}$:
\begin{equation}
{\cal F}^j_{(a)} = \left[\delta_k^j (p^l p_l) - p_k p^j \right] {\cal W}^k_{(a)} \,.
\label{F2}
\end{equation}
The second construction contains antisymmetric object ${\cal V}^{jl}_{(a)}$
\begin{equation}
{\cal F}^j_{(a)} = {\cal V}^{jl}_{(a)} p_l \,, \quad {\cal V}^{jl}_{(a)} = - {\cal V}^{lj}_{(a)} \,,
\label{F3}
\end{equation}
and can be in principle rewritten in the first form.  In our context one can classify all forces orthogonal to the particle momentum using the so-called Effective Field Theory \cite{EFT}, which considers zero, first, etc. order terms in derivatives of vector, pseudoscalar, electromagnetic fields, as well as, in derivatives of the space-time metric.

\subsubsection{Zero-order force terms}

We know only one construction, which does not contain derivatives, namely
\begin{equation}
{\cal F}^i_{(S)} = \frac{\lambda_{(S)}}{m_a c}[\delta^i_k (p^l p_l) - p^ip_k] \ U^k \ \left[1 + \nu_{(S)}\cos{\phi} \right]  \,.
\label{F50}
\end{equation}
It is based on the aether velocity four-vector $U^k$, contains even periodic function of the axion field, and includes two phenomenologically introduced scalar functions $\lambda_{(S)}$ and $\nu_{(S)}$. We assume that these scalars can depend on the expansion scalar of the aether flow, $\Theta = \nabla_k U^k$, but can be the constant also. This force belongs to the class of non-gyroscopic ones, since the divergency
\begin{equation}
\frac{\partial}{\partial p^i} {\cal F}^i_{(S)}= -3 \frac{\lambda_{(S)}}{m_a c} (p_k U^k) \left[1 + \nu_{(S)}\cos{\phi} \right]
\label{F51}
\end{equation}
is non-vanishing. The index $(S)$ emphasizes that in classical hydrodynamics there exists the so-called Stokes friction force, which is proportional to the difference between the velocities of particle and of the hydrodynamic flow. Here the aether flow  plays the hydrodynamic role, and the force vanishes, when the particle four-velocity $\frac{p^j}{m_{(a)}c}$ coincides with the aether velocity $U^j$.  When $|\nu_{(S)}|>1$, the multiplier $\left[1 {+} \nu_{(S)}\cos{\phi} \right]$ behaves as  axionic switch, providing the force to be of the friction or antifriction type depending on the state of the axion field.

\subsubsection{First-order force terms}

This subclass contains two forces of the rotational type based on derivatives of the vector field
\begin{equation}
{\cal F}^i_{(R)} = \frac{\lambda_{(R)}}{m_a c}\left[\nabla^i U^k - \nabla^k U^i \right] \ p_k \ \left[1 + \nu_{(R)}\cos{\phi} \right] \,,
\label{F61}
\end{equation}
\begin{equation}
{\cal F}^i_{(*R)} = \frac{\lambda_{(*R)}}{m_a c} \sin{\phi} \ \epsilon^{ikmn}\nabla_m U_n \ p_k\  \left[1 + \nu_{(*R)}\cos{\phi} \right]  \,.
\label{F62}
\end{equation}
Both forces are of the gyroscopic type, since
\begin{equation}
\frac{\partial}{\partial p^i} {\cal F}^i_{(R)}= 0 \,, \quad \frac{\partial}{\partial p^i} {\cal F}^i_{(*R)}= 0 \,.
\label{F63}
\end{equation}

There are two gradient-type forces, which contain the gradient of the axion field $\nabla_k \phi$. The first one
\begin{equation}
{\cal F}^i_{(G)} = \frac{\lambda_{(G)}}{m_a c} \ [g^{ik} (p^l p_l) - p^i p^k]  \ \nabla_k \phi \ \sin{\phi}  \left[1 + \nu_{(G)}\cos{\phi} \right]
\label{F73}
\end{equation}
is non-gyroscopic with
\begin{equation}
\frac{\partial}{\partial p^i} {\cal F}^i_{(G)}= -3 \frac{\lambda_{(G)}}{m_a c} \ \sin{\phi} \ p^k \ \nabla_k \phi \ \left[1 + \nu_{(G)}\cos{\phi} \right] \,.
\label{F731}
\end{equation}
The second gradient-type force
\begin{equation}
{\cal F}^i_{(*G)} = \frac{\lambda_{(*G)}}{m_a c} \epsilon^{ikls} U_l \ p_s \ \nabla_k \phi \ \left[1 + \nu_{(*G)}\cos{\phi} \right]
\label{F81}
\end{equation}
contains the Levi-Civita (pseudo)tensor, the aether velocity four-vector, and  is of the gyroscopic type.

There are two gyroscopic forces, which can be indicated as generalizations of the Lorentz force. The first force of this type contains the Maxwell tensor linear in derivatives of the electromagnetic potential
\begin{equation}
{\cal F}^i_{(L)} = \frac{e_a}{m_a c^2}F^i_{\cdot k} p^k \left[1 + \nu_{(L)}\cos{\phi} \right]  \,.
\label{F91}
\end{equation}
When $\nu_{(L)}=0$ it is, clearly, the classical Lorentz force. The second generalization can be constructed as follows:
\begin{equation}
{\cal F}^i_{(*L)} = \frac{e^*_a}{m_a c^2}\sin{\phi} F^{*i}_{\ \cdot k} p^k  \left[1 + \nu_{(*L)}\cos{\phi} \right]   \,.
\label{F92}
\end{equation}
It contains the dual Maxwell tensor and some hypothetical dual charge $e^*_a$. We have to stress, that the scalar functions   $\lambda_{(R)}$ and $\nu_{(R)}$, $\lambda_{(*R)}$ and $\nu_{(*R)}$, $\lambda_{(G)}$ and $\nu_{(G)}$, $\lambda_{(*G)}$ and $\nu_{(*G)}$, $\nu_{(L)}$ and $\nu_{(*L)}$ can be considered as functions of the expansion scalar $\Theta$, or as constants.

\subsubsection{On the second-order force terms}

When we deal with the forces containing two derivatives, there are a lot of variants of the force modeling. First, we can use twice the derivative of the vector field, $(\nabla_k U_m) (\nabla_l U_n)$; second, the terms quadratic in the gradient of the axion field $(\nabla_k \phi)(\nabla_n \phi)$ can appear; third, the forces containing the product of Maxwell fields $F_{mn} F_{pq}$ are also admissible. Then we have to list all the cross-terms containing $(\nabla_k U_m) \nabla_n \phi$, $(\nabla_k U_m) F_{ls}$, $\nabla_k \phi F_{mn}$. Finally, the terms with second derivatives can appear, say, $\nabla_k \nabla_m U_n$, $\nabla_k \nabla_n \phi$, $\nabla_k F_{mn}$. We keep in mind these terms, but we do not consider them in the context of linear electrodynamics and field theory of the second order in derivatives. Only one specific class of forces attracts the interest in our theory; it is the class of the so-called tidal forces, linear in curvature.
In order to formulate these tidal forces in the so-called non-minimal way, we introduce the three-parameter tensor of non-minimal susceptibility
\begin{equation}
{\cal R}_{ikmn} = \frac12 q_1 R \left(g_{im}g_{kn}-g_{in}g_{km}\right) + q_2 \left[ R_{im} g_{kn} - R_{in} g_{km} + R_{kn} g_{im} - R_{km} g_{in} \right] + q_3 R_{ikmn} \,,
\label{Fr921}
\end{equation}
based on the Riemann tensor $R_{ikmn}$, Ricci tensor $R_{km}$ and Ricci scalar $R$. Convolutions of this tensor give two supplementary obgects
\begin{equation}
{\cal R}_{im} = \left(\frac32 q_1 +q_2 \right) R g_{im} + \left(2q_2 + q_3 \right) R_{im} \,, \quad {\cal R} = R (6q_1+6q_2+q_3) \,.
\label{Frr92}
\end{equation}
Also, there exist left-dual and right-dual susceptibility tensors
\begin{equation}
^*{\cal R}^i_{ \cdot kmn} = \frac12 \epsilon^i_{\cdot kpq} \ {\cal R}^{pq}_{\ \ mn} \,, \quad {\cal R}^{*i}_{\  \cdot kmn} = \frac12 {\cal R}^{i}_{\cdot k pq} \epsilon^{pq}_{\ \ mn}  \,.
\label{FD1}
\end{equation}
With these tensors of non-minimal susceptibility we can construct three tidal forces of the second order in derivative of the metric.
The first one is based on the tensor ${\cal R}_{ikmn}$
\begin{equation}
{\cal F}^i_{(T)} = \frac{\lambda_{(T)}}{m_a c} {\cal R}^i_{\cdot kmn} p^k U^m p^n \left[1 + \nu_{(T)}\cos{\phi} \right]\,,
\label{FD2}
\end{equation}
and has non-vanishing divergency
\begin{equation}
\frac{\partial}{\partial p^i} {\cal F}^i_{(T)}= - \frac{\lambda_{(T)}}{m_a c} {\cal R}_{km} p^k U^m  \left[1 + \nu_{(T)}\cos{\phi} \right]\,.
\label{FD3}
\end{equation}
The second and third tidal forces are of the gyroscopic type, since they are constructed using the dual tensors
\begin{equation}
{\cal F}^i_{(T*)} = \frac{\lambda_{(T*)}}{m_a c} \sin{\phi} \ {\cal R}^{*i}_{\ \ \cdot kmn} p^k U^m p^n \left[1 + \nu_{(T*)}\cos{\phi} \right]\,, \quad \frac{\partial}{\partial p^i} {\cal F}^i_{(T*)}= 0 \,,
\label{FD4}
\end{equation}
\begin{equation}
{\cal F}^i_{(*T)} = \frac{\lambda_{(*T)}}{m_a c} \sin{\phi} \ ^*{\cal R}^{i}_{\ \cdot kmn} p^k U^m p^n \left[1 + \nu_{(*T)}\cos{\phi} \right] \,, \quad \frac{\partial}{\partial p^i} {\cal F}^i_{(*T)}= 0 \,.
\label{FD5}
\end{equation}
Again, we have to repeat, that the scalar functions   $\lambda_{(T)}$ and $\nu_{(T)}$, $\lambda_{(*T)}$ and $\nu_{(*T)}$, $\lambda_{(T*)}$ and $\nu_{(T*)}$ can be considered as functions of the expansion scalar $\Theta$, or as constants.

\section{Equilibrium states in axionically - aetherically active plasma}

\subsection{General concept and basic definitions }

We follow the standard definition of the equilibrium state, which requires that the entropy production scalar of the closed system vanishes. For elastic collisions this requirement is satisfied, when the collision integrals vanish, ${\cal J}_{(a)(b)}=0$, and the distribution functions are of the J\"uttner - Chernikov form \cite{RKT}:
\begin{equation}
f^{(\rm eq)}_{(a)} = A_{(a)} e^{- \xi_k p^k } \,.
\label{eq1}
\end{equation}
The scalar function $A_{(a)}$ is connected with the normalization of the distribution function. The vector field $\xi^k$ has no the sort index and has to be the time-like one, i.e., $\xi_k \xi^k >0$. The norm of this four-vector is usually associated with the temperature of the system, $T(x)$, $\frac{c}{k_{B} T(x)} = \sqrt{\xi_k \xi^k}$, where $k_{B}$ is the Boltzmann constant. The unit time-like four-vector $V^j = \xi^j (\xi_k \xi^k)^{- \frac12}$ defines the macroscopic velocity of the kinetic system. The quantities
$A_{(a)}$ and $\xi_k$ should be found from the equation
\begin{equation}
p^j \frac{\partial }{\partial x^j}\left[ \log{A_{(a)}}\right] - \frac12 p^j p^k \left[\nabla_j \xi_k + \nabla_k \xi_j \right] = m_{(a)} c \ \xi_j {\cal F}_{(a)}^j - m_{(a)} c \ \frac{\partial}{\partial p^j} {\cal F}_{(a)}^j \,.
\label{eq2}
\end{equation}
It is well known that for the simple massive gas, when ${\cal F}_{(a)}^j=0$, the equilibrium state is admissible, when $\nabla_j \xi_k {+} \nabla_k \xi_j {=}0$ and thus $\xi^k$ has to be time-like Killing vector. When one deals with the massless gas, $m_{(a)}=0$, the necessary condition for the equilibrium state takes the form $\nabla_j \xi_k {+} \nabla_k \xi_j = 2\Psi g_{jk}$, and thus requires that $\xi^k$  to be the conformal Killing vector. In both cases the functions $A_{(a)}$ convert into constants.
When ${\cal F}_{(a)}^j \neq 0$, we have to decompose the forces into the series with respect to momentum $p^k$ and to solve the set of equations obtained in the first, second, etc. orders of this decomposition.

Let us discuss this problem for the forces listed above.
Taking into account the terms quadratic in the particle momenta, we obtain the equation for the vector $\xi_j$
\begin{equation}
\nabla_{(m} \xi_{n)} = - g_{mn} \left[{\cal H}_{(S)} \xi^k U_k + {\cal H}_{(G)} \sin{\phi} \ \xi^k \nabla_k \phi \right] +
\label{EQ11}
\end{equation}
$$
+ \left[{\cal H}_{(S)} \xi_{(m} U_{n)} + {\cal H}_{(G)} \sin{\phi} \ \xi_{(m} \nabla_{n)} \phi \right] +
$$
$$
+ \xi^j U^s \left[{\cal H}_{(T)} {\cal R}_{j (mn)s} + {\cal H}_{(*T)} \sin{\phi} \  ^*{\cal R}_{j(mn)s} + {\cal H}_{(T*)} \sin{\phi} {\cal R} \ ^{*}_{j(mn)s} \right]
 \,.
$$
Here the parentheses denote symmetrization, and definitions are made of the following type
\begin{equation}
{\cal H}_{(S)} = \lambda_{(S)} \left[1+ \nu_{(S)} \cos{\phi} \right] \,, ...\,, {\cal H}_{(T*)} = \lambda_{(T*)} \left[1+ \nu_{(T*)} \cos{\phi} \right] \,.
\label{EQ13}
\end{equation}
The equation (\ref{EQ11}) can be rewritten in more compact form
\begin{equation}
\nabla_{(m} \xi_{n)} = - g_{mn} \xi^k {\cal Q}_k + \xi_{(m}  {\cal Q}_{n)} +
\xi^{(j} U^{s)}  {\cal W}_{jmns} \,,
\label{EQ11y}
\end{equation}
where the following auxiliary quantities are introduced:
\begin{equation}
{\cal Q}_k =  \left[{\cal H}_{(S)} U_k + {\cal H}_{(G)} \sin{\phi} \ \nabla_k \phi \right] \,,
\label{EQ18}
\end{equation}
\begin{equation}
{\cal W}_{jmns}= \left[{\cal H}_{(T)} {\cal R}_{j (mn)s} + {\cal H}_{(*T)} \sin{\phi} \  ^*{\cal R}_{j(mn)s} + {\cal H}_{(T*)} \sin{\phi} {\cal R} \ ^{*}_{j(mn)s} \right]
 \,.
\label{EQ19}
\end{equation}
Convolutions of (\ref{EQ11y}) with $g^{mn}$ and $\xi^m \xi^n$ give, respectively
\begin{equation}
\nabla_m \xi^m  = - 3 \xi^k {\cal Q}_k - {\cal H}_{(T)} {\cal R}_{js} \xi^{j} U^{s}   \,, \quad
\xi^m  \nabla_m \left[\xi_n \xi^n \right]  = 0 \,.
\label{EQ21}
\end{equation}
Collecting the terms linear in $p^j$ we obtain equations for the quantities $A_{(a)}$
\begin{equation}
\nabla_k \left(\log{A_{(a)}} \right) = {\cal H}_{(R)} \xi^j \left(\nabla_j U_k - \nabla_k U_j \right)+
{\cal H}_{(*R)} \sin{\phi} \xi^j \epsilon_{jkmn} \nabla^m U^n
+ {\cal H}_{(*G)} \xi^j \epsilon_{jlks} U^l \nabla^s \phi +
\label{EQ17}
\end{equation}
$$
+ \frac{e_{(a)}}{c} \left[1{+} \nu_{(L)} \cos{\phi} \right] \xi^j F_{jk}
+ \frac{e^*_{(a)}}{c} \sin{\phi} \left[1{+} \nu_{(*L)} \cos{\phi} \right] \xi^j F^*_{jk}
+ 3U_k {\cal H}_{(S)} + 3 {\cal H}_{(G)} \sin{\phi} \nabla_k \phi
+ {\cal H}_{(T)} {\cal R}_{km} U^m
\,.
$$
Convolution of (\ref{EQ17}) with $\xi^k$ yields
\begin{equation}
\xi^k \nabla_k \left(\log{A_{(a)}} \right) = 3 \xi^k U_k {\cal H}_{(S)} + 3 {\cal H}_{(G)} \sin{\phi} \xi^k \nabla_k \phi
+ {\cal H}_{(T)} {\cal R}_{km} \xi^k U^m
\,.
\label{EQ31}
\end{equation}

\subsection{Special ansatz}

Our ansatz is that the four-vector $\xi^j$ is proportional to the aether velocity $U^j$, i.e., $\xi^j = \frac{c U^j}{k_{B} T}$.
This ansatz is compatible with equations (\ref{EQ11y}), when
\begin{equation}
\nabla_{(m} U_{n)} - \frac{1}{T} U_{(m} \nabla_{n)}T = - g_{mn} U^k {\cal Q}_k + U_{(m}  {\cal Q}_{n)} +
U^j U^{s}  {\cal W}_{jmns} \,.
\label{EQ119}
\end{equation}
 Also, we have to take into account that
 \begin{equation}
 U^m \nabla_m T=0 \,, \quad \nabla_m U^m =  - 3 U^k {\cal Q}_k - {\cal H}_{(T)} {\cal R}_{js} U^{j} U^{s} \,.
 \label{EQ86}
\end{equation}
Let us check that this ansatz works in the model of homogeneous isotropic cosmological model.

\subsection{Example of application}

Let us consider the isotropic homogeneous cosmological model of the FLRW type with the  metric
\begin{equation}
 ds^2 = dt^2 - a^2(t) \left[(dx^1)^2 + (dx^2)^2 +(dx^3)^2 \right] \,.
 \label{metric}
\end{equation}
It is well known that the space-time of this type does not admit the existence of a time-like Killing vector, thus the standard equilibrium of massive particles is impossible. However, the dynamic aether could provide the existence of equilibrium of a new type.
For illustration we consider the simple situation, when, first, all quantities depend on time only; second, the axion field is in the equilibrium state at the minimum of the potential indicated by the number $n$, i.e., $\phi = 2 \pi n$; third, ${\cal H}_{(T)} = 0$.

Now the aether velocity is of the form $U^j = \delta^j_0$, and the covariant derivative takes very simple form
\begin{equation}
\nabla_m U_n = \frac13 \Theta \Delta_{mn} \,, \quad \Delta_{mn}= g_{mn} - U_m U_n \,.
\label{metric2}
\end{equation}
The equations (\ref{EQ119}) transforms into
\begin{equation}
\frac13 \Theta \Delta_{mn} - \frac{\dot{T}}{T} U_{m} U_n  = - {\cal H}_{(S)} \Delta_{mn} \,,
\label{EQ1193}
\end{equation}
 thus providing the relationships
 \begin{equation}
 \dot{T}=0 \,, \quad H =  - {\cal H}_{(S)}\,.
 \label{EQr86}
 \end{equation}
 Here we used the standard definition of the Hubble function $H = \frac{\dot{a}}{a} = \frac13 \Theta$.
In fact, this relationship predetermines the structure of the multiplier $\lambda_{(S)}(H)$:
 \begin{equation}
 \lambda_{(S)}(H) = - \frac{H}{[1+ \nu_{(S)}]} \ \Rightarrow \ \lambda_{(S)} = - \tilde{\lambda}_{(S)} H \,, \quad  \tilde{\lambda}_{(S)} \equiv \frac{1}{1+ \nu_{(S)}} \,.
 \label{EQ76}
\end{equation}
In other words, the equilibrium state in plasma exists, the macroscopic velocity of plasma flow coincides with the aether velocity and the equilibrium temperature does not depend on time. In this context the multiplier $\lambda_{(S)}(H)$ happens to be linear in the Hubble function, and $\nu_{(S)}$ is constant.
The coefficient $A_{(a)}$ satisfies now the equations
\begin{equation}
\nabla_k \left(\log{A_{(a)}} \right) = 3U_k {\cal H}_{(S)} \,,
\label{EQp17}
\end{equation}
which reduce to one equation $\frac{\dot{A}_{(a)}}{A_{(a)}} = -3 H$. The solution to this  equation is $A_{(a)}(t)=A_{(a)}(t_0) \left[\frac{a(t_0)}{a(t)} \right]^3$. Since the normalization coefficient $A_{(a)}$ is proportional to the particle number density ${\cal N}_{(a)}$, the established law simply shows that, when the Universe expands, the particle number decreases as ${\cal N}_{(a)}(t) = {\cal N}_{(a)}(t_0) \left[\frac{a(t_0)}{a(t)} \right]^3$.

\section{Outlook}

1. The formalism of the kinetic theory of the relativistic multi-component axionically - aetherically active plasma is presented. The self-consistent set of coupled master equations contains five subsets: first, the kinetic equations for the distribution functions (\ref{KE1}); second, the extended Jacobson's equations for the unit vector field, describing the velocity of the dynamic aether (\ref{AEq}); third, the extended electrodynamic equations (\ref{T1}); fourth, the extended equations for the axion field (\ref{24}); fifth, the gravity field equations (\ref{25}) - (\ref{T91}).

2. The classification of the extended forces, which act on the plasma particles in the electromagnetic, aetheric, axionic and gravitational fields, is suggested in the framework of the Effective Field Theory up to the second order in derivatives (see Subsection IIE). This classification is supplemented by phenomenological modeling of the additional source terms, which appear in the equations of the electromagnetic, axion and vector fields due to the plasma back-reaction.

3. The concept of extended equilibrium in axionically - aetherically active plasma is suggested and is tested using the example of isotropic homogeneous cosmological model.


\section*{References}


\begin{thebibliography}{99}

\bibitem{RKT} S.R. de Groot, W.A. van Leeuwen, Ch.G, van Weert. Relativistic kinetic theory. (North-Holland Publishing Company. Amsterdam - New York- Oxford), 1980.

\bibitem{P1}
Pitaevskii L.P. and Lifshitz E.M.  Physical Kinetics (Butterworth-Heinenann Ltd, Oxford), 1981.

\bibitem{P2} A. Piel.  Plasma Physics. An Introduction to Laboratory, Space and Fusion Plasmas (Springer Verlag, Berlin, Heidelberg), 2010.

\bibitem{P3} V.P. Silin and A.A. Rukhadze.  Electromagnetic Properties of Plasma and Plasmalike Media. (3rd Edition, Publishing House Librokom, Moscow), 2013.

\bibitem{Jackson}
J.D. Jackson. Classical Electrodynamics. (John Wiley and Sons, USA), 1999.

\bibitem{Hehl} F.W. Hehl and Yu.N. Obukhov. Foundations of Classical
Electrodynamics: Charge, Flux, and Metric (Birkh\"auser:
Boston, USA), 2003.

\bibitem{Sikivie} P. Sikivie. Experimental tests of the "invisible" axion. Phys. Rev. Lett. 1983, {\bf 51}, 1415--1417.


\bibitem{Wilczek1987} F. Wilczek. Two applications of axion electrodynamics. Phys. Rev. Lett. 1987, {\bf 58}, 1799--1802.

\bibitem{PQ}
R.D. Peccei and  H.R. Quinn.  CP conservation in the presence of instantons. Phys. Rev. Lett. 1977, {\bf 38}, 1440--1443.

\bibitem{W1} S. Weinberg. A new light boson? Phys. Rev. Lett. 1978, {\bf 40}, 223--226.

\bibitem{W2} F. Wilczek Problem of strong P and T invariance in the presence of instantons. Phys. Rev. Lett. 1978, 
{\bf 40}, 279--282.

\bibitem{B0} A.B. Balakin, R.K. Muharlyamov  and A.E. Zayats. Electromagnetic waves in an axion-active relativistic plasma non-minimally coupled to gravity. The European Physical Journal C – Particles and Fields. 2013, {\bf 73}, 2647.

\bibitem{B1} A.B. Balakin, R.K. Muharlyamov  and A.E. Zayats. Nonminimal Einstein-Maxwell-Vlasov-axion model. Classical and Quantum Gravity. 2014, {\bf 31}, 025005.

\bibitem{B2} A.B. Balakin, R.K. Muharlyamov  and A.E. Zayats. Axion-induced oscillations of cooperative electric field in a cosmic magneto-active plasma. The European Physical Journal D. 2014, {\bf 68}, 159.

\bibitem{B3} A.B. Balakin and D.E. Groshev. Polarization and stratification of axionically active plasma in a dyon magnetosphere.  Phys. Rev. D 2019, {\bf 99}, 023006.

\bibitem{J1} T. Jacobson and D. Mattingly. Gravity with a dynamical
preferred frame. Phys. Rev. D 2001, {\bf 64}, 024028.

\bibitem{J2} T. Jacobson and D. Mattingly. Einstein-aether waves. Phys. Rev. D 2004, {\bf 70}, 024003.

\bibitem{J3}
 C. Heinicke, P. Baekler and F.W. Hehl. Einstein-aether
theory, violation of Lorentz invariance, and metric-affine gravity. Phys. Rev. D 2005, {\bf 72}, 025012.

\bibitem{BaLe} A.B. Balakin and J.P.S. Lemos. Einstein-aether theory with a Maxwell field: General formalism. Annals of Physics. 2014, {\bf 350}, 454-484.

\bibitem{AA} A.B. Balakin. Axionic extension of the Einstein-aether theory. Phys. Rev. D. 2016, {\bf 94}, 024021.

\bibitem{anti1} W. Zimdahl and A.B. Balakin. Inflation in a self-interacting gas universe. Phys. Rev. D 1998, {\bf 58}, 063503.

\bibitem{anti2} W. Zimdahl, D.J. Schwarz, A.B. Balakin and D. Pavon. Cosmic anti-friction and accelerated expansion. Phys. Rev. D 2001, {\bf 64}, 063501.

\bibitem{anti3} W. Zimdahl and A.B. Balakin. Cosmological thermodynamics and deflationary gas universe. Phys. Rev. D 2001, {\bf 63}, 023507.

\bibitem{anti4} A.B. Balakin, D. Pavon, D.J. Schwarz and W. Zimdahl. Curvature force and dark energy.
New J. Phys. 2003, {\bf 5}, 85.

\bibitem{SH1} A.B. Balakin and A.F. Shakirzyanov. Axionic extension of the Einstein-aether theory: How does dynamic aether regulate the state of axionic dark matter? Physics of the Dark Universe 2019, {\bf 24}, 100283.


\bibitem{SH2}  A.B. Balakin and A.F. Shakirzyanov. Isotropic Cosmological Model with Aetherically Active Axionic Dark Matter.  Universe 2024, {\bf 10}, 74.

\bibitem{SH3} A.B. Balakin and A.F. Shakirzyanov. The extended Einstein-Maxwell-aether-axion theory: Effective metric as an instrument of the aetheric control over the axion dynamics.  Gravitation and Cosmology 2024, {\bf 30}, 57-67.

\bibitem{WTN} Wei-Tou Ni. Equivalence principles and electromagnetism. Phys. Rev. Lett. 1977, {\bf 38}, 301-304.

\bibitem{EFT} C.P. Burgess.  Introduction to effective field theory. Ann. Rev. Nucl. Part. Sci. 2007, {\bf 57}, 329.




\end{thebibliography}
\end{document}